\documentclass[aps,prl,superscriptaddress,unsortedaddress,twocolumn]{revtex4}
\usepackage{amsmath,epsfig,mathrsfs,graphicx,amssymb,slashed}

\begin{document}
\title{Justification of the canonical quantization of the Josephson effect}
\author{Krzysztof Pomorski}
\email{Krzysztof.Pomorski@fuw.edu.pl}
\affiliation{Faculty of Physics, University of Warsaw, ul. Pasteura 5, PL02-093 Warsaw, Poland}
\author{Adam Bednorz}
\email{Adam.Bednorz@fuw.edu.pl}
\affiliation{Faculty of Physics, University of Warsaw, ul. Pasteura 5, PL02-093 Warsaw, Poland}
\date{\today}
\begin{abstract}
Quantum devices based on Josephson effect in superconductors are usually described by a Hamiltonian obtained by commonly used canonical quantization. However, this recipe has not been yet rigorously justified.
We show that this approach is indeed correct in certain range of parameters. We find the condition of the validity of such quantization and the lowest corrections to the Josephson energy.
\end{abstract}
\maketitle
The Josephson effect \cite{jos} is one of most fundamental phenomena occurring in physical systems with macroscopic quantum effects. When biased by external capacitance, inductance or current source, it becomes the basis for the
description of many superconducting quantum devices such as Cooper pair boxes, flux and phase superconducting qubits \cite{phasqu,dev}.
It has been demonstrated that such structures can be promising elements for quantum computing \cite{comp} and violation of local
realism (Bell test, although only local at present) \cite{mart}. The detailed theoretical description of such a system
becomes more and more necessary for its highly complex and demanding experimental applications.

The simplest, and to date most popular and phenomenological description of the quantum devices based on Josephson effect is derived from current-phase relation,
phase-voltage relation.
The quantum properties of Josephson junction are extracted by using conjugated observables for example between number particles (charge) and phase  \cite{cano}. The energy stored in capacitor decides whether we deal with Cooper pair box, flux or phase qubit.
On the other hand, the whole system is already originally quantum, with superconductor (conventional, $s$-wave) described at least by standard Bardeen-Cooper-Schrieffer (BCS) theory \cite{bcs,bdg} so there should be no need to quantize it again. The canonical quantization procedure is just a shortcut between the full quantum description and the approximated model \cite{wid}. Certainly one should be able to justify this procedure and find the range of its validity. A partial justification has been developed 30 years ago \cite{ambe}, where the effective action of the Josephson junction has been found and -- in path integral picture of quantum mechanics -- the canonically quantized Hamiltonian is revealed in the tunneling limit. However, that approach does not give an answer to a very relevant question  of the range of validity, except qualitative statement that phase fluctuations should be smaller than the superconductor energy gap. The point is that tunneling limit (low transmission through the junctions) is not a sufficient condition as both transmission, number of modes and capacitance play the role. If one simply takes tunneling limit at constant capacitance then also whole useful quantum properties are lost. Phase fluctuations are indeed intuitively of the order of capacitance energy but one has to settle it quantitatively.

In this work, we derive the approximate Hamiltonian (corresponding to the one obtained by canonical quantization) straight from BCS Hamiltonian, using standard perturbative approach with both transmission \emph{and} capacitance treated as small parameters. By introducing phase-dependent eigenstates, the formalism takes much analogy from adiabatic and nonadiabatic transitions. We find that in the lowest order, which can be called \emph{adiabatic approximation} \cite{adiab}, one indeed finds the desired simple effective Hamiltonian. The next order modifies Josephson energy and ratio of this modification gives the range of applicability of the adiabatic approximation. Interestingly, the modification contains an \emph{infinite} term, which is fortunately phase-independent and so it has no physical relevance, and appears also in the non-superconducting case. The phase-dependent part can be calculated analytically also for large transmission, but it is rather technical so the details are presented in Supplemental Material \cite{supl}.
Higher order terms become rather complicated, but do not contain anymore infinity.

\begin{figure}
\includegraphics[scale=.5]{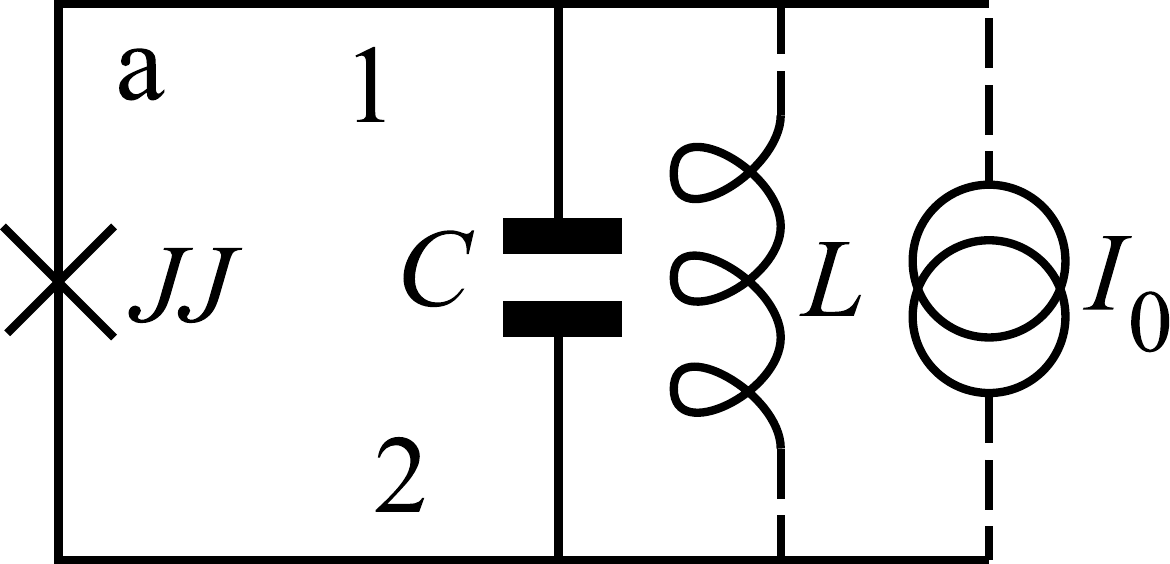}
\includegraphics[scale=.5]{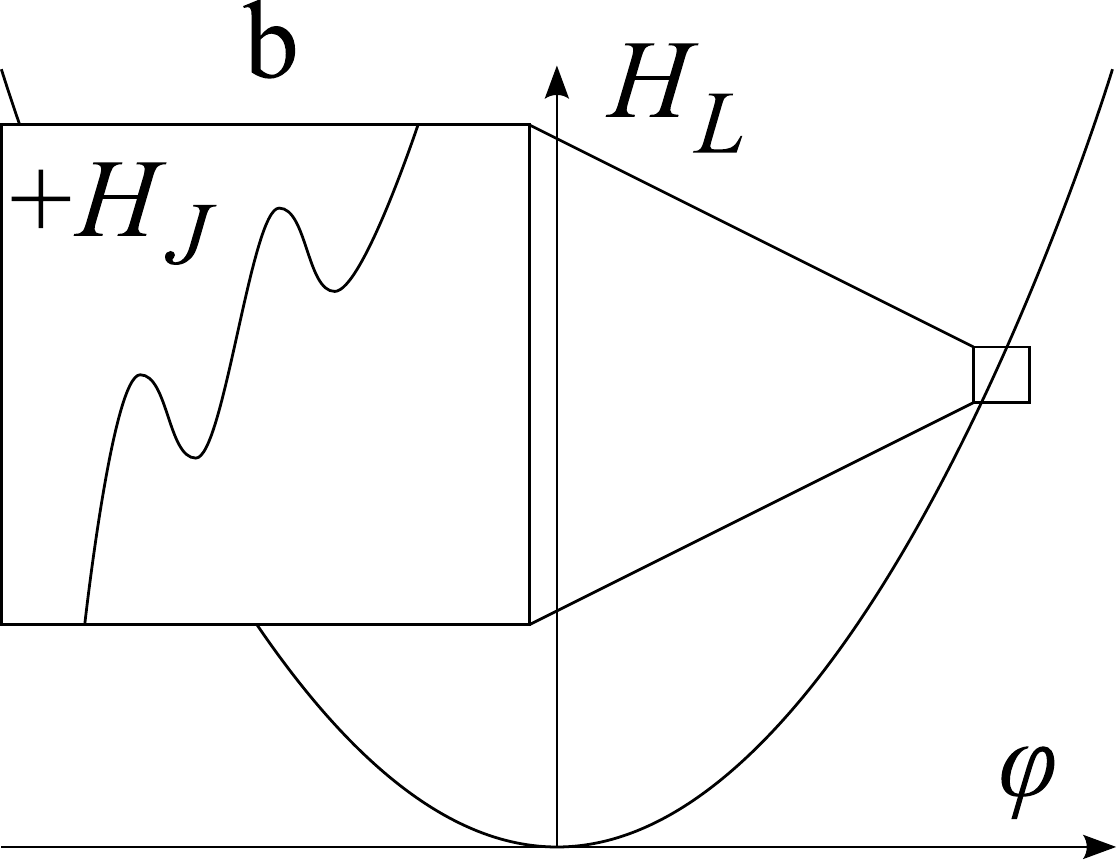}
\includegraphics[scale=.5]{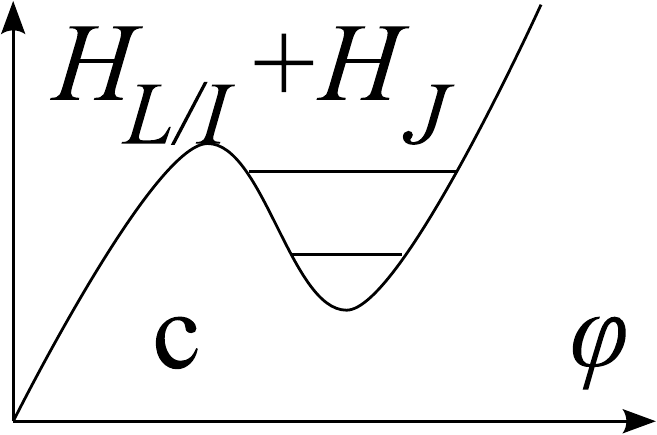}
\caption{(a)The effective electric circuit of the Josephson junction (JJ) biased by capacitance $C$, inductance $L$ and/or current $I_0$ with distinguished parts 1,2.  (b) The energy $H_{L/I}+H_J$ with the quantum operation range in the inset. (c) Quantum energy levels in the potential landscape}\label{phq}
\end{figure}

Let us begin with the sketch of the common canonical quantization of an electric circuit with Josephson effect \cite{cano}.
The system depicted in Fig. \ref{phq}, consists of the Josephson junction (JJ), effective capacitance $C$ and  bias inductance $L$ or current $I_0$.
The total Hamiltonian will be the sum of the contributions of all these elements. We have $H_C=Q^2/2C=CV^2/2$,
where $Q$ is the charge and $V$ is the voltage between $1$ and $2$. The voltage can be represented by \emph{phase} $\phi$
change ratio $V=\hbar\dot\phi/e$ ($2\phi=\Phi_1-\Phi_2$ for the respective superconducting phases $\Phi_{1,2}$). Analogously, $H_L=LI^2/2=(\hbar\phi/e)^2/2L$ with the current $I=\dot{Q}$. Alternatively, one can consider current bias $I_0$ giving the term $H_I=I_0\hbar \phi/e$ ($I_0$ is then the current between $1$ and $2$ in absence of other elements). The \emph{classical} picture leads to relation $\dot{Q}=I=e\partial H/\hbar\partial\phi$. On the other hand the Josephson current is calculated as $I_J=I_1\sin(2\phi)$ \cite{jos}. Using the above classical equation, this leads to Josephson energy $H_J=-e\hbar I_1\cos(2\phi)/2$. Now, the canonical quantization means that we take the Hamiltonian $H=H_C+H_{L/I}+H_J$ as a function of conjugate variables $\phi$ and $Q$ and replace everything by their quantum counterparts, including conversion of Poisson bracket into commutator between $\phi$ and $Q$. It essentially means $Q\to e\partial/i\partial\phi$.
We stress that the canonical quantization of the $LIC$ part ($H_C+H_{L/I}$) is already well justified by quadratic form of the electromagnetic action \cite{elec}, whose Lagrangian density has  the form $\mathcal L=\epsilon_0|\vec{E}|^2/2-|\vec{B}|^2/2\mu_0-\rho V+\vec{A}\cdot\vec{j}$, where $\vec{E}=-\nabla V-\partial_t \vec{A}$,$\vec{B}=\nabla\times\vec{A}$,$V$,$\vec{A}$,$\rho$,$\vec{j}$,$\epsilon_0$,$\mu_0$ denote electric and magnetic field, scalar and vector potential, charge density and current, electric and magnetic permittivity, respectively, and linear relations between $V$ and $\vec{E}$, $Q$ and $\rho$, $I$ and $\vec{j}$. Certainly, one has to assume ideal models of capacitance and inductance, which will surely break down  at high frequencies or voltages (e.g. leakage currents).
Nevertheless, the useful range of quantum operation parameters is well within the regime of validity of the assumed model of the $LIC$ part.
One usually considers the classical turning points defined by $E\sim H_L+H_J$, where $H_L$ can be approximated by $e^2\phi_0\phi/Le^2$ with the exact crossing point $\phi_0$, equivalent to current bias term $H_I$, see Fig. \ref{phq}b. Therefore the most suspicious part of this procedure is not the quantization of $LI$ or $C$ terms but JJ. On one hand, the Josephson effect is crucial for the appearance of quantum levels in the potential landscape, see Fig.\ref{phq}c. On the other hand, the effect is no longer quadratic and originally derived straight from BCS theory plus tunneling. It is intuitive that going back-and-forth between quantum and classical description is not always accurate and may lead to losing some corrections. To find this accuracy is the main goal of this work.

We shall model the full Hamiltonian of the superconductor ($s$-wave) by the standard bulk BCS part written in the Bogoliubov-de Gennes framework \cite{bdg} $H_{BCS}$, plus tunneling term $H_T$. We shall use the relation between second and first quantized Hamiltonian
$H=\Psi^\dag h\Psi$ with anticommutation $\{\Psi^\dag,\Psi\}=I$ and $\{\Psi,\Psi\}=0$ (here $\Psi$ can be multi-component). The first quantization Hamiltonian $h=h_{BCS}+h_T$ is defined in a one-dimensional bispinor basis
$x$, $(1,2)\times(e,h)$, where $x$ is the (real) position in the superconductor $1$ or $2$ and $e/h$ denotes electron/hole. In the standard BCS approximation (small superconducting gap), for a single channel, we have
\begin{equation}
h_{BCS}=
\begin{pmatrix}
-i\partial_x& \Delta\\
\Delta& i\partial_x\end{pmatrix}_{eh},h_{eT}=-\tau\delta(x)\begin{pmatrix}
0&e^{i\phi}\\
e^{-i\phi}&0\end{pmatrix}_{12}\label{hhh}
\end{equation}
and $h_{hT}=-h_{eT}(\phi\to-\phi)$ where $h_{BCS}$ is the same in subspaces $1$ and $2$ while $h_T$ has different phase in electron and hole subspace.
Here $\Delta$ is the superconducting gap and position is represented in time units divided by $\hbar$. Normal scattering matrix is then
\begin{equation}
S_e=\begin{pmatrix}
r&it e^{i\phi}\\
it e^{-i\phi}&r\end{pmatrix}\label{smat}
\end{equation}
with $t=\sin(\tau/\hbar)$ and $r=\sqrt{1-t^2}$,$S_h=S_e^T$. One can add the overall phase $e^{i\alpha/\hbar}$ to the scattering matrix by modifying $h_T\to h_T+\alpha\delta(x)$ (accounting interface properties) but it will not change any of our results and hence we can safely disregard it. We stress that putting $\phi$ into the tunneling term instead of varying phases at $\Delta$ in the BCS part is only matter of the gauge choice. In our gauge the phase $\phi$ corresponds simply to the vector potential across the junction $\int_1^2e\vec{A}\cdot d\vec{r}/\hbar$ so that superconducting phases remain constant all the time.

\begin{figure}
\includegraphics[scale=.5]{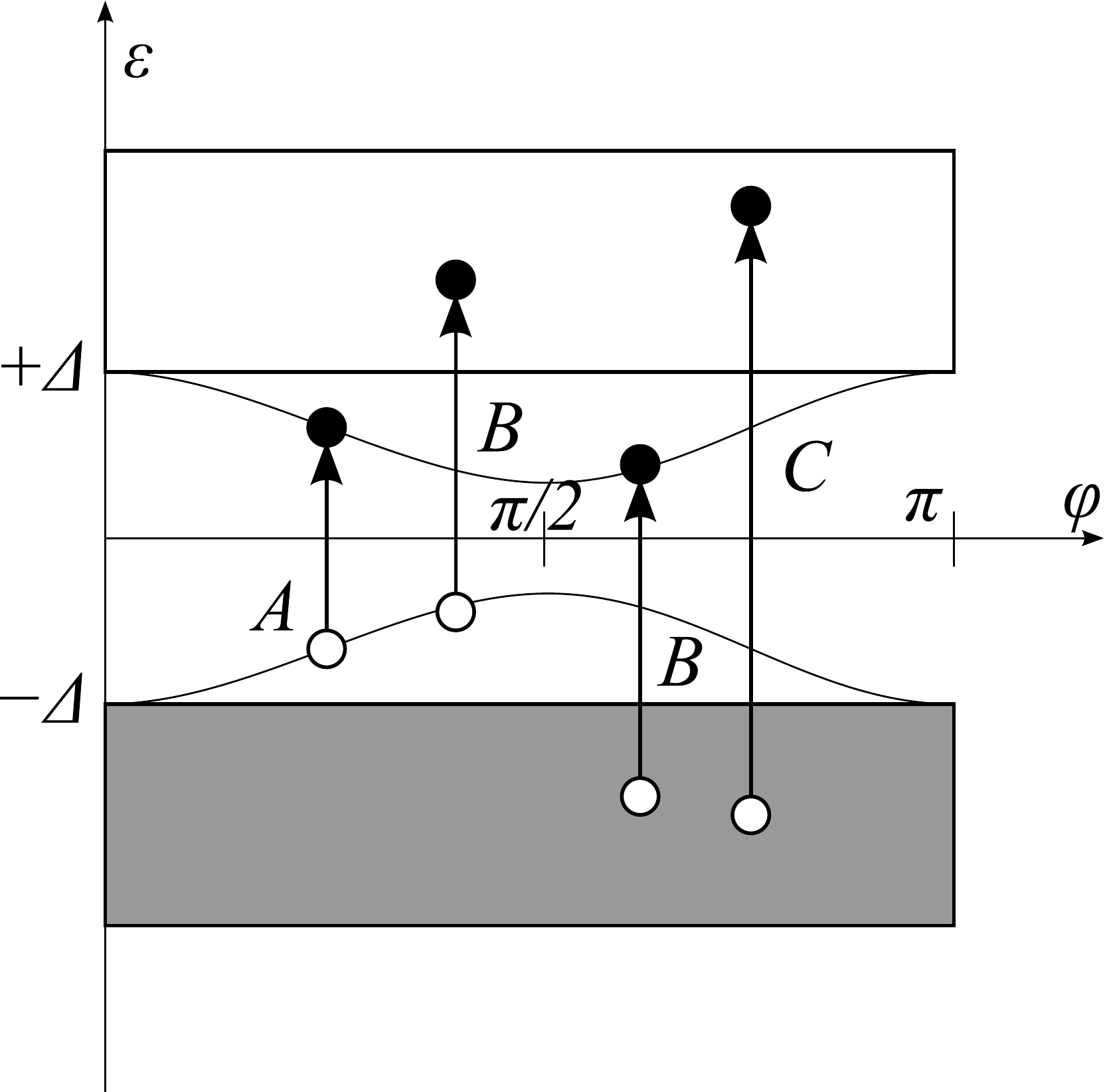}
\caption{Energy spectrum of JJ, with the continuum parts above $+\Delta$ and below $-\Delta$ and two phase-dependent ABSs in the gap.  Possible excitations of the ground state: (A) between ABSs, (B) between an ABS and continuum, (C) between lower and upper continuum.}\label{abslev}
\end{figure}

Now one can calculate the ground energy (we focus on zero temperature case) of the whole Hamiltonian assuming Fermi level at $0$ (all states with negative energy $\epsilon$ are occupied). The Josephson effect follows from  the fact that this energy is \emph{phase-dependent}.
The spectrum of $h_{BCS}$ consists of continuum positive and negative parts separated by the gap $|\epsilon|\geq \Delta$ \cite{bcs}. Introducing $h_T$ modifies the wavefunction of these (now scattering) levels but not their energies. However, $h_T$ allows for new, so-called Andreev bound states (ABS) \cite{absref}, with energies inside the gap $|\epsilon|<\Delta$ and hence localized at the junction point ($x=0$).
These energies (see detailed calculation in Supplemental Material \cite{supl} which is anyway quite straightforward \cite{abscal}) are equal
$\epsilon_\pm=\pm\Delta\sqrt{1-t^2\sin^2\phi}$ and, for small $t$, $\simeq \pm\Delta\mp\Delta t^2\sin^2\phi/2$, see Fig. \ref{abslev}. Taking $H_J\simeq \epsilon_-$,
and summing over independent channels, $\sum_t$, as usually JJ contains many,
we have the desired canonical quantized scheme. Note also that the Josephson current can be justified as $I=\partial e\partial H_J/\hbar\partial\phi$ by means of full counting statistics \cite{fcs}, where $\langle e^{i\chi Q(t)}\rangle
=\langle e^{itH(\phi-\hbar\chi/2e)/\hbar} e^{-itH(\phi+\hbar\chi/2e)/\hbar}\rangle$ in the ground state, which follows from Keldysh formalism \cite{kel}, and is in fact equivalent to Josephson's original derivation \cite{jos}.

The above presented analysis, although leads to the desired result, is incomplete. It is not only the energy that changes with the phase $\phi$ but also the whole wavefunction. One has to ensure that correction from these changes is sufficiently small in the quantum operation regime, defined by $e^2/C\sim \Delta\sum_tt^2$. The argument that wavefunction changes become small for $t\to 0$ (i.e. $\ll e^2/\Delta C$) is insufficient because all relevant quantum properties (energy levels) disappear in this limit. To find rigorously the range of validity of the canonical approximation we have to calculate the effect of wavefunction changes.

We shall treat the problem by standard perturbative approach, keeping $t^2$ and $e^2/C$ as the small parameters \emph{at the same order}. Without $H_{T/C/L/I}$ we  are left with the ground state of $H_{BCS}$ Hamiltonian, denoted by $|g\rangle$, degenerated by all possible values of $\phi$ and energy $E_g=0$ (reference value). All excited states will be denoted by $|e\rangle$ with energies $E_e>0$. Furthermore, we can find ground and excited states for  $H_{T}$ included, but now they become in general $\phi$-dependent, $|g_\phi\rangle$ with energy $E_{g\phi}$ and $|e_\phi\rangle$ with energy $E_{e\phi}$.
For a given $\phi$ all such states are orthonormal and form a complete basis, but states for \emph{different} $\phi$ are not necessarily orthonormal.

The full eigenproblem reads
$
E|\psi\rangle=H|\psi\rangle
$,
where the \emph{zero order} state is $|g\rangle$ with energy $0$. Since $H_T$ is small, all states $|g_\phi\rangle$ will also reduce to $|g\rangle$ in zero order. Therefore we can write an ansatz for $|\psi\rangle$ in the  form
$
|\psi\rangle=\int d\phi(\psi(\phi)|g_\phi\rangle+\sum_e\psi_e(\phi)|e_\phi\rangle)
$,
where the latter sum is of higher order. Plugging this form into the eigenproblem we get
\begin{eqnarray}
&&E'\psi|g_\phi\rangle+\sum_e E'\psi_e|e_\phi\rangle=\left(E_{g\phi}-\frac{e^2}{2C}\partial_\phi^2\right)\psi|g_\phi\rangle
\nonumber\\
&&+\sum_e\left(E_{e\phi}-\frac{e^2}{2C}\partial_\phi^2\right)\psi_e|e_\phi\rangle\label{eee}
\end{eqnarray}
with $E'=E-H_{L/I}$.
Sandwiching it with $\langle g_\phi|$ we get
\begin{eqnarray}
&&E'\psi=E_{g\phi}\psi-\frac{e^2}{2C}\partial_\phi^2\psi-(\partial_\phi\psi) (e^2/C)\langle g_\phi|\partial_\phi|g_\phi\rangle\nonumber\\
&&-\psi(e^2/2C)\langle g_\phi|\partial_\phi^2|g_\phi\rangle
-\frac{e^2}{2C}\sum_e\langle g_\phi|\partial_\phi^2\psi_e|e_\phi\rangle.\label{eee1}
\end{eqnarray}
The factor $\langle g_\phi|\partial_\phi|g_\phi\rangle$ is an analogue to the differential \emph{Berry phase} \cite{berry}, which can be arbitrarily chosen. For our purpose, it is convenient to assume that it is zero. Only \emph{relative} Berry phase for excited states would matter but only at high perturbation order. In the lowest order, we obtain
\begin{equation}
E'\psi\simeq E_{g\phi}\psi-\frac{e^2}{2C}\partial_\phi^2\psi-\psi(e^2/2C)\langle g_\phi|\partial_\phi^2|g_\phi\rangle.
\end{equation}
The first two terms can be interpreted as adiabatic approximation \cite{adiab}, while the last one is a non-adiabatic correction to
$E_{g\phi}=\epsilon_-$ which is the main term we are here interested in.
The next order term is a bit more technical and contains $\partial_\phi$, see \cite{supl}.
Even higher terms get very complicated and very likely already obscured by other effects so there is no practical reason to discuss them.

\begin{figure}
\includegraphics[scale=.5,angle=270]{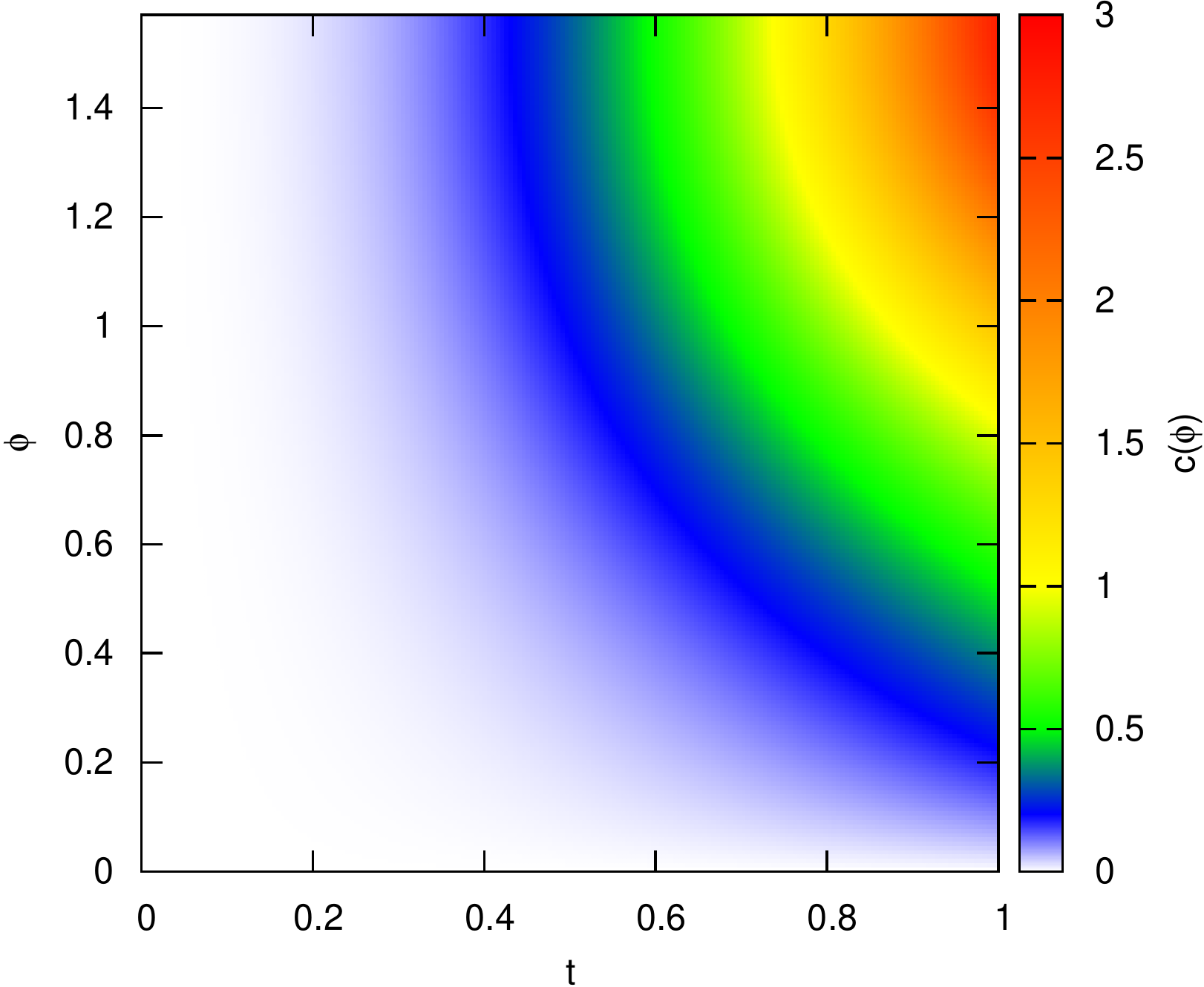}
\caption{The exact dependence of $c(\phi)$  on $\phi\in[0,\pi/2]$ and $t$ referenced at $\phi=0$, $c$ is even in $\phi$ and has the period $\pi$.}\label{wykr}
\end{figure}

The term $\langle g_\phi|\partial_\phi^2|g_\phi\rangle$ can be evaluated inserting identity between derivatives,
$
\langle g_\phi|\partial_\phi^2|g_\phi\rangle=\langle g_\phi|\partial_\phi|g_\phi\rangle\langle g_\phi\partial_\phi|g_\phi\rangle+\sum_e\langle g_\phi|\partial_\phi|e_\phi\rangle\langle e_\phi|\partial_\phi|g_\phi\rangle
$
since we assume zero differential ground Berry phase, and from orthogonality between $g$ and $e$ we get
$
\langle g_\phi|\partial_\phi^2|g_\phi\rangle=-
\sum_e|\langle g_\phi|\partial_\phi|e_\phi\rangle|^2=-\sum_e|\langle e_\phi|\partial_\phi|g_\phi\rangle|^2
$
The final form of the equation with  lowest correction is
\begin{equation}
E'\psi\simeq E_{g\phi}\psi-\frac{e^2}{2C}\partial_\phi^2\psi+\psi(e^2/2C)c(\phi)
\label{eee2}
\end{equation}
Here $c(\phi)=\sum_e |f_{e\phi}|^2$, with $f_{e\phi}=\langle e_\phi|\partial_\phi|g_\phi\rangle$,  is the (dimensionless) prefactor of the main correction to the Josephson energy $E_{g\phi}$.
As we consider the zero temperature case, in the ground state all single-particle states with \emph{negative} energies are fully occupied while \emph{positive} energies are fully empty. The relevant excited states (for the terms
$\langle g|\partial|e\rangle$) are those with a \emph{single} excitation, namely one state with positive energy gets fully occupied at the cost of full emptying one of the negative energy states see Fig.\ref{abslev}.
Note also that for many independent channels both terms have to be summed over channels, indexed by their transmission $t$, namely
$\sum_t E_{g\phi}$ and $\sum_t c_t(\phi)$.
One can write down $c(\phi)$ in terms of first-quantized ($oc$-cupied and $em$-pty) single particles states
\begin{equation}
c(\phi)=\sum_{oc,em}|\langle \psi_{em}|\partial_\phi|\psi_{oc}\rangle|^2.\label{ccc}
\end{equation}
One can also use the adiabatic identity
\begin{equation}
\langle g_\phi|\partial_\phi|e_\phi\rangle=\frac{\langle g_\phi|(\partial_\phi H_T)|e_\phi\rangle}{E_{e\phi}-E_{g\phi}}
\end{equation}
because only $H_T$ depends on $\phi$. This is already sufficient to find the lowest contribution in $t$. In this approximation states $g$ and $e$ become $\phi$-independent. Putting $E_g=0$ as a reference we are left with excited states represented by $|\theta_+,\theta_-\rangle$ where $\theta_\pm$ represent the single-particle state in the positive/negative spectrum.
Since $H_T$ transfers between $1$ and $2$ the states $\theta$ are in different superconductors. In momentum space we have
$k=\Delta\sinh\theta$ and $|\theta_\pm\rangle=(2\cosh\theta)^{-1/2}(e^{\pm\theta_\pm/2},\pm e^{\mp\theta_\pm/2})^T$ in the $eh$ basis,
with $\epsilon_\pm(\theta)=\pm\Delta\cosh\theta$ and $E_e=\epsilon_+(\theta_+)-\epsilon_-(\theta_-)$. In this approximation
\begin{equation}
c(\phi)\simeq 2t^2\int \frac{d\theta_-d\theta_+}{(4\pi)^2}\frac{|e^{(\theta_+-\theta_-)/2+i\phi}-e^{(\theta_--\theta_+)/2-i\phi}|^2}
{(\cosh\theta_++\cosh\theta_-)^2}.
\end{equation}
By introducing variables $2s=\theta_++\theta_-$ and $2w=\theta_+-\theta_-$ we get
\begin{equation}
c(\phi)\simeq t^2\int \frac{dsdw}{(2\pi)^2}\frac{\sinh^2w+\sin^2\phi}
{\cosh^2 s\cosh^2 w}.
\end{equation}
The integral contains an infinite but phase-independent part limited by energy cutoff $c_{div}\sim (t/\pi)^2\ln(\epsilon_{\max}/\Delta)$. Interestingly this term appears already in the normal case, and can be related to Fermi edge singularity \cite{edge}. There are no more any infinities in higher order terms, and so this infinity has no physical relevance. The phase-dependent part is fortunately finite and equal $t^2\sin^2\phi/\pi^2$.
This is our main result. The use of canonical quantization is valid only if the correction to Josephson energy is small,
i.e. $e^2/2C\ll\Delta$. This confirms the intuition \cite{ambe} that the correction is small at phase variation $\sim E_C/\hbar$ ($E_C=e^2/2C$) smaller than $\Delta/\hbar$. Remember that the quantum operation regime leads also to the condition $t^2\ll 1$ or $\sum_t t^2\ll 1$ in the multichannel case. Note also,
that in the lowest order only $\Delta$ appearing in Josephson energy gets renormalized to $\Delta'=\Delta+e^2/2C\pi^2$; there are no qualitative
differences. The qualitative change appears in higher order terms $\sim t^4$ but also in the adiabatic ground energy $E_{g\phi}$, so the whole description gets anyway complicated \cite{tink}. Note that higher order terms will also contain corrections to $\Delta$ near the tunneling point (also in $E_{g\phi}$). Nevertheless, one can calculate exact $c(\phi)$ analytically for all $t$ (not only small) assuming constant (bulk) $\Delta$,
which is quite lengthy and requires considering all possible excitations shown in Fig.\ref{abslev}, see details in the Supplemental Material \cite{supl}. Such excitations are relevant also in other applications of ABS \cite{abse}. Here we present this general result only
graphically in Fig. \ref{wykr}.

In conclusion, we have found rigorous condition of validity of canonical quantization of the Josephson effect in presence of capacitance, $e^2/C\ll \Delta$, showing that the lowest correction simply renormalizes Josephson energy, and constructing a general procedure to find higher corrections. The lowest correction may be useful in fine-tuning of quantum energy levels.

The thank W. Belzig and W. Bardyszewski for helpful remarks and motivation.

\onecolumngrid
\newpage
\vspace{5cm}

\begin{center}
\large\bf Supplemental Material
\end{center}
\vspace{1cm}

\section{A. Systematic perturbative approach to Josephson effect with capacitance}
\renewcommand{\theequation}{A.\arabic{equation}}
\setcounter{equation}{0}

We shall show how to get at least two lowest order corrections to the adiabatic canonically quantized Hamiltonian, outlining the general procedure. We start from (\ref{eee}) and (\ref{eee1}) from the main text. Sandwiching (\ref{eee}) with $\langle e_\phi$, in the lowest order, we get
\begin{equation}
E_{e\phi}\psi_e\simeq (e^2/2C)\langle e_\phi|\partial_\phi^2\psi|g_\phi\rangle
\end{equation}
Plugging it into the last term of (\ref{eee1}) we get the  desired higher order term.
Note also that in the lowest order
$
\langle g_\phi|\partial_\phi^2\psi_e|e_\phi\rangle\simeq2(\partial_\phi\psi_e)\langle g_\phi|\partial_\phi|e_\phi\rangle
$ and
$
\langle e_\phi|\partial_\phi^2\psi|g_\phi\rangle\simeq 2(\partial_\phi\psi)\langle e_\phi|\partial_\phi|g_\phi\rangle
$.
The final form of the equation with two lowest corrections is
\begin{equation}
E'\psi\simeq E_{g\phi}\psi-\frac{e^2}{2C}\partial_\phi^2\psi+\psi(e^2/2C)\sum_e|f_{e\phi}|^2
+\frac{e^4}{C^2}\sum_e f^\ast_{e\phi}\partial_\phi(E^{-1}_{e\phi}\partial_\phi(\psi f_{e\phi}))\label{eee3}
\end{equation}
with $f_{e\phi}=\langle e_\phi|\partial_\phi|g_\phi\rangle$.

\section{B. Analytic calculation of $c(\phi)$.}
\renewcommand{\theequation}{B.\arabic{equation}}
\setcounter{equation}{0}

\subsection{Scattering matrix}

\begin{figure}
\includegraphics[scale=0.5]{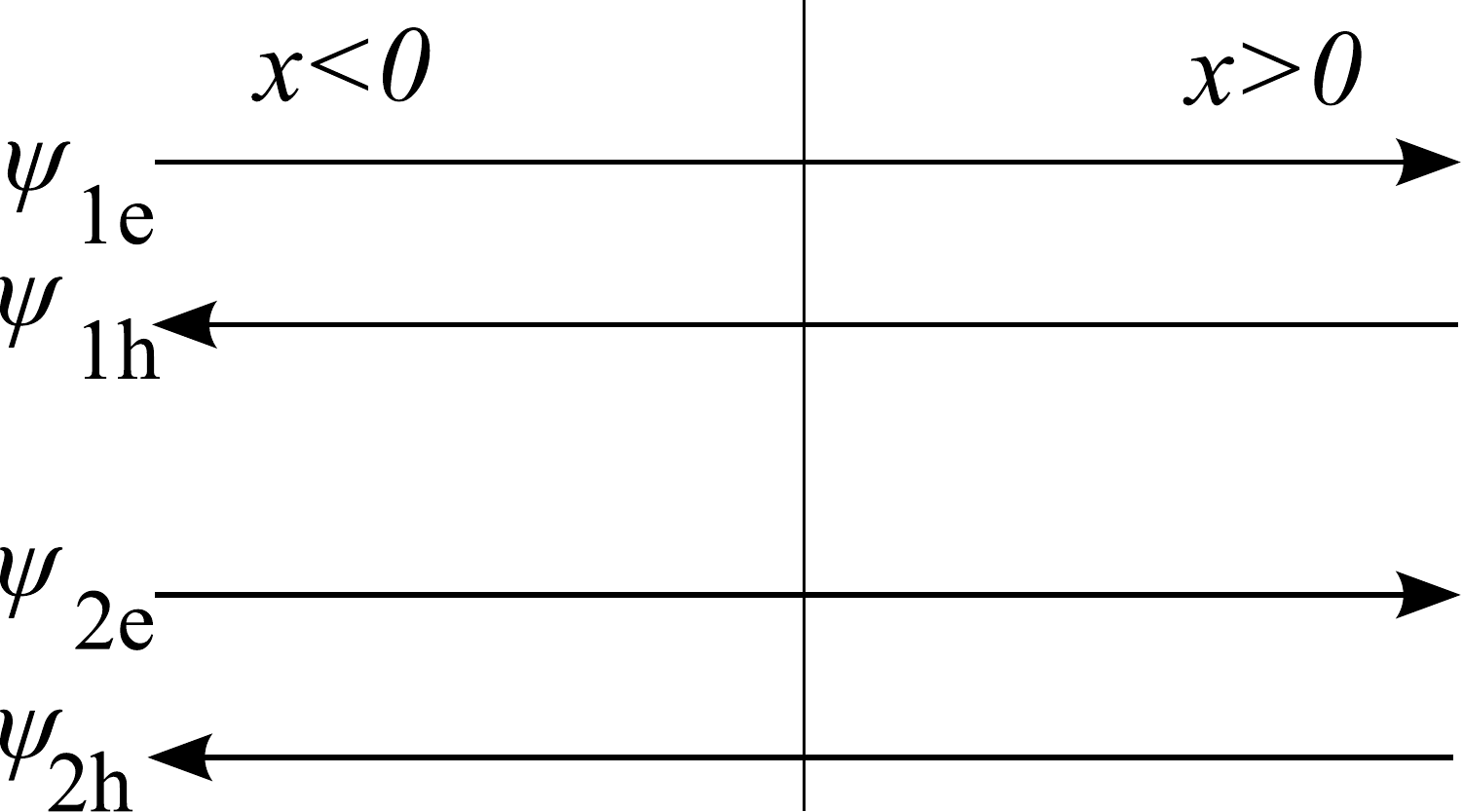}
\caption{The propagation of $e$-lectrons and $h$-oles. We shall use the notation $R/L$ for the modes at $x>/<0$}
\label{prop}
\end{figure}

\begin{figure}
\includegraphics[scale=0.5]{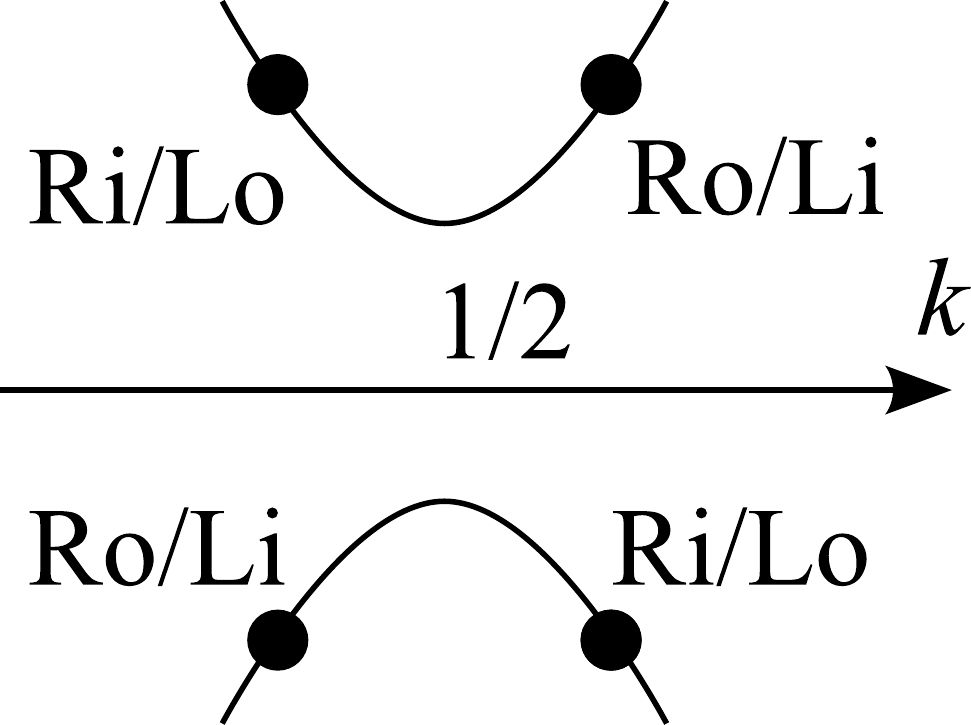}
\caption{The $k$ dependence of direction of BCS eigenstates, based on group velocity $\partial\epsilon/\partial k$.}
\label{inout}
\end{figure}

To operate on the continuum states we need to find their scattering matrix. Note that the natural directions of electron/hole propagation as in Fig. \ref{prop}, may be different from in/out-going scattering eigenstates of $h_{BCS}$
because the direction depends on the sign of group velocity $\partial \epsilon/\partial k$ as depicted in Fig. \ref{inout}

We have 4 incoming and 4 outgoing modes depicted below, indexed by $nAa$, where
$n=1,2$ denotes the superconductor $A=R,L$ (right/left) denotes the side of wave $(R\equiv x=0+,L\equiv x=0-)$ and
$a=i,o$ denotes $i$-ncoming and $o$-utgoing mode,see also Fig.\ref{inout}.

The scattering matrix connects the modes amplitudes
\begin{equation}
A_o=
\begin{pmatrix}
A_{Ro}\\
A_{Lo}\end{pmatrix}
=
\begin{pmatrix}
A_{1Ro}\\
A_{2Ro}\\
A_{1Lo}\\
A_{2Lo}\end{pmatrix}
=S
\begin{pmatrix}
A_{1Li}\\
A_{2Li}\\
A_{1Ri}\\
A_{2Ri}\end{pmatrix}
=S\begin{pmatrix}
A_{Li}\\
A_{Ri}\end{pmatrix}=SA_i
\end{equation}

We take $\theta>0$. For  energies $\pm\Delta\cosh\theta$ we  take $\theta_{Ro}=\theta_{Li}=\pm\theta$ and
$\theta_{Ri}=\theta_{Lo}=\mp\theta$.
Then the boundary condition in normal scattering matrix (\ref{smat}) leads to
\begin{eqnarray}
&&e^{\theta/2}A_{Ro}+e^{-\theta/2}A_{Ri}=S_e(e^{\theta/2}A_{Li}+e^{-\theta/2}A_{Lo})\nonumber\\
&&e^{-\theta/2}A_{Ro}+e^{\theta/2}A_{Ri}=S_h(e^{-\theta/2}A_{Li}+e^{\theta/2}A_{Lo})
\end{eqnarray}
equivalent to
\begin{equation}
\begin{pmatrix}
e^{\theta/2}&-e^{-\theta/2}S_e\\
e^{-\theta/2}&-e^{\theta/2}S_h
\end{pmatrix}
\begin{pmatrix}
A_{Ro}\\
A_{Lo}\\
\end{pmatrix}
=\begin{pmatrix}
e^{\theta/2}S_e&-e^{-\theta/2}\\
e^{-\theta/2}S_h&-e^{\theta/2}
\end{pmatrix}
\begin{pmatrix}
A_{Li}\\
A_{Ri}\\
\end{pmatrix}
\end{equation}
so
\begin{equation}
S=\begin{pmatrix}
e^{\theta/2}&-e^{-\theta/2}S_e\\
e^{-\theta/2}&-e^{\theta/2}S_h
\end{pmatrix}^{-1}
\begin{pmatrix}
e^{\theta/2}S_e&-e^{-\theta/2}\\
e^{-\theta/2}S_h&-e^{\theta/2}
\end{pmatrix}
\end{equation}
We have to solve
\begin{equation}
\begin{pmatrix}
e^{\theta/2}&-e^{-\theta/2}S_e\\
e^{-\theta/2}&-e^{\theta/2}S_h
\end{pmatrix}
\begin{pmatrix}
a&b\\
c&d
\end{pmatrix}=1
\end{equation}
which is equivalent to
\begin{eqnarray}
&&e^{\theta/2}a-e^{-\theta/2}S_ec=1,\:e^{\theta/2}b=e^{-\theta/2}S_ed\nonumber\\
&&e^{-\theta/2}a=e^{\theta/2}S_hc,\:e^{-\theta/2}b-e^{\theta/2}S_hd=1
\end{eqnarray}
Substituting $a=e^\theta S_hc$ and $b=e^{-\theta}S_ed$ we get
$
(e^{\theta}S_h-e^{-\theta}S_e)c=e^{-\theta/2}$ and $(e^{-\theta}S_e-e^\theta S_h)d=e^{\theta/2}
$

Let us focus on the matrix
$
W=e^\theta S_h-e^{-\theta}S_e=(e^{\theta} \bar{S}_h-e^{-\theta}\bar{S}_e)
$
which appears in
\begin{equation}
\begin{pmatrix}
e^{\theta/2}&-e^{-\theta/2}S_e\\
e^{-\theta/2}&-e^{\theta/2}S_h
\end{pmatrix}^{-1}=\begin{pmatrix}
e^{\theta/2}S_h&-e^{-\theta/2}S_e\\
e^{-\theta/2}&-e^{\theta/2}\end{pmatrix}W^{-1}
\end{equation}
We have
\begin{equation}
e^{\theta} \bar{S}_h-e^{-\theta}\bar{S}_e=2\begin{pmatrix}
r\sinh\theta& it\sinh(\theta-i\phi)\\
it\sinh(\theta+i\phi)&r\sinh\theta
\end{pmatrix}
\end{equation}
The determinant is equal $4M$, where $M=|\sinh(\theta+i\phi)|^2=\sinh^2\theta+t^2\sin^2\phi$.
Then
$
W^{-1}=(4M)^{-1}(e^{\theta} \bar{S}_h^\dag-e^{-\theta}\bar{S}_e^\dag)
$
Finally
\begin{equation}
4MS=
\begin{pmatrix}
e^{\theta/2}\bar{S}_h&-e^{-\theta/2}\bar{S}_e\\
e^{-\theta/2}&-e^{\theta/2}
\end{pmatrix}(e^{\theta} \bar{S}_h^\dag-e^{-\theta}\bar{S}_e^\dag)
\begin{pmatrix}
e^{\theta/2}\bar{S}_e&-e^{-\theta/2}\\
e^{-\theta/2}\bar{S}_h&-e^{\theta/2}
\end{pmatrix}
\end{equation}
or explicitly
\begin{equation}
\begin{pmatrix}
r\sinh^2\theta&-t\sinh\theta\sin(\phi-i\theta)&-t^2\sin\phi\sin(\phi-i\theta)&-rt\sinh\theta\sin\phi\\
t\sinh\theta\sin(\phi+i\theta)&r\sinh^2\theta&rt\sinh\theta\sin\phi&-t^2\sin\phi\sin(\phi+i\theta)\\
-t^2\sin\phi\sin(\phi+i\theta)&-rt\sinh\theta\sin\phi&r\sinh^2\theta
&-t\sinh\theta\sin(\phi+i\theta)\\
rt\sinh\theta\sin\phi&-t^2\sin\phi\sin(\phi-i\theta)&t\sin\theta\sinh(\phi-i\theta)&r\sinh^2\theta
\end{pmatrix}\label{sss}
\end{equation}

\subsection{Andreev bound states}
Apart from the continuum states, there are evanescent subgap solutions of the equation $\epsilon=h$ for $h=h_{BCS}+h_T$
given by (\ref{hhh}). For evanescent modes $e^{\kappa x}\psi$, denoting $\kappa=\Delta\sin\beta$, $|\beta|\leq \pi/2$
we get eigenvalues $\epsilon_\pm=\pm \Delta\cos\beta$ with eigenvectors
$$
\psi_\pm=2^{-1/2}\begin{pmatrix}
e^{\mp i\beta/2}\\
\pm e^{\pm i\beta/2}\end{pmatrix}_{eh}
$$
Certainly these modes can exist only in the vicinity of the junction point $x=0$ forming bound states.
The Andreev bound states are formed by left ($L$ for $x<0$) and right ($R$ for $x>0$) evanescent modes.
Let us take $\beta_R=-\beta_L=\beta<0$ for the  $\epsilon_+$ and $\beta_R=-\beta_L=-\beta<0$
for $\epsilon_-$. We use the boundary condition at $x=0$ to get
\begin{eqnarray}
&&e^{-i\beta/2}A_R=e^{i\beta/2}S_eA_L\nonumber\\
&&e^{i\beta/2}A_R=e^{-i\beta/2}S_hA_L\label{abse}
\end{eqnarray}
for the scattering matrices given by (\ref{smat}).
Hence
$A_R=e^{i\beta}S_eA_L=e^{-i\beta}S_hA_L
$
The condition for existence of solution is
\begin{equation}
0=\det(e^{i\beta}S_e-e^{-i\beta}S_h)=4\det
\begin{pmatrix}
ir\sin\beta&-t\sin(\beta+\phi)\\
-t\sin(\beta-\phi)&ir\sin\beta
\end{pmatrix}
\end{equation}
The last determinant is equal
$-r^2\sin^2\beta-t^2\sin^2\beta\cos^2\phi+t^2\cos^2\beta\sin^2\phi$
which is $-\sin^2\beta+t^2\sin^2\phi$. Since it must vanish, we get the condition
\begin{equation}
\sin\beta=\pm t\sin\phi\label{abs}
\end{equation}
the amplitudes are found by
$
ir\sin\beta A_{L1}=t\sin(\beta+\phi)A_{L2}
$.
Plugging (\ref{abs}) we get
$
\pm ir\sin\phi A_{L1}=(\pm t\sin\phi\cos\phi+\cos\beta\sin\phi)A_{L2}
$
which lead to
$
\pm irA_{L1}=(\cos\beta\pm t\cos\phi)A_{L2}
$.
Note that $\cos\beta=(1-t^2\sin^2\phi)^{1/2}=(r^2+t^2\cos^2\phi)^{1/2}>|t\cos\phi|$
and
$
(\cos\beta+ t\cos\phi)(\cos\beta-t\cos\phi)=\cos^2\beta-t^2\cos^2\phi=r^2
$
so we have
$
\pm i(\cos\beta\mp t\cos\phi)^{1/2}A_{L1}=(\cos\beta\pm t\cos\phi)^{1/2}A_{L2}
$.
The normalized solution reads
\begin{equation}
A_L=2^{-1/2}\begin{pmatrix}
(1\pm t\cos\phi/\cos\beta)^{1/2}\\
\pm i(1\mp t\cos\phi/\cos\beta)^{1/2}
\end{pmatrix}
\label{absl}
\end{equation}
Rewriting equations (\ref{abse}) in the form
\begin{eqnarray}
&&e^{-i\beta}A_R=S_eA_L\nonumber\\
&&e^{i\beta}A_R=S_hA_L\label{abse2}
\end{eqnarray}
and subtracting the second from the first equation we get
\begin{equation}
-2i\sin\beta A_R=(S_e-S_h)A_L
=2\begin{pmatrix}
0&-t\sin\phi\\
t\sin\phi&0\end{pmatrix}A_L
\end{equation}
and finally, plugging (\ref{abs}) and (\ref{absl}) we get
\begin{equation}
A_R=\begin{pmatrix}
(1\mp t\cos\phi/\cos\beta)^{1/2}\\
\pm i(1\pm t\cos\phi/\cos\beta)^{1/2}
\end{pmatrix}
\end{equation}
The final wavefunctions have to be normalized by multiplying by $|t\sin\phi|^{1/2}$.

\subsection{Berry phase}

The Berry phase itself will be irrelevant for the lowest orders of perturbative approach.
Nevertheless, we shall need it because of usefulness in other calculations.
We ask what happens when the eigenbasis is parameter-dependent. The parameter is $\phi$ in our case.
For the nondegenerate eigenvalues the differential Berry phase is defined by
$
\langle\psi|\partial_\phi|\psi\rangle=i\gamma
$
where $\gamma$ is real because
$
0=\partial_\phi\langle\psi|\psi\rangle=i(\gamma-\gamma^\ast)
$
For degenerate case we collect all the states of the same energy. Assume (it will be anyway our case)
that degeneracy is not lifted by changing $\phi$ (except some singular points). We construct matrix $\gamma$
with
$
\langle\psi_m|\partial_\phi|\psi_n\rangle=i\gamma_{mn}
$
Note that
$
0=\partial_\phi\langle\psi_m|\psi_n\rangle=i(\gamma_{mn}-\gamma_{nm}^\ast)
$
which shows that $\gamma$ is Hermitian. We can find individual Berry phases by diagonalizing $\gamma$. Then
eigenvalues will be differential Berry phases.

Let us calculate first the Berry phase for Andreev bound states.
The derivative of spatial part is real so it cannot give any contribution to Berry phase.
Acting on $e^{\pm i\beta/2}$ the net result is also zero as both signs appear with equal weight.
Finally acting on $A_{R/L}$ we get real numbers, so again they cannot give any Berry phase. The conclusion is that
the phase convention we used in construction of Andreev bound states representatives gives zero Berry phase.

For scattering states we have the total state combined from incoming wave and scattered outgoing wave
$
\psi_a=\psi_{ia}+\sum_c\psi_{oc}S_{ca}
$
where $a,c=R/L,1/2$. Only $S$ can depend on $\phi$ so the matrix $\gamma$ is  given by
\begin{equation}
i\gamma_{ba}=\left(\langle\psi_{ib}|+S^\dag_{bd}\sum_d\langle\psi_{od}|\right)
\sum_c|\psi_{oc}\rangle \partial_\phi S_{ca}
\end{equation}
The overlap of $Ri/Ro$ and  $Li/Lo$ is $\sim i/\sinh\theta$, so it does not contain any $\delta$ term
(the $i/o$ states have opposite momentum) so the term vanishes in the continuum limit (infinite length of the junction).
Therefore
$
i\gamma=S^\dag\partial_\phi S/2
$
because the outgoing waves are halves of full waves.
The final calculation gives $2M\gamma=$

\begin{equation}
t\sinh\theta
\begin{pmatrix}
-t\cosh\theta&ir\cosh(\theta+i\phi)&0&
i\cos\phi\\
-ir\cosh(\theta-i\phi)&t\cosh\theta&
-i\cos\phi&0\\
0&i\cos\phi&t\cosh\theta&ir\cosh(\theta-i\phi)\\
-i\cos\phi&0&-ir\cosh(\theta+i\phi)&-t\cosh\theta
\end{pmatrix}
\end{equation}

We shall also need
$
i\tilde\gamma=S\partial_\phi S^\dag/2
$
which is $2M\tilde\gamma=$\
\begin{equation}t\sinh\theta
\begin{pmatrix}
-t\cosh\theta&-ir\cosh(\theta+i\phi)&0&
-i\cos\phi\\ir\cosh(\theta-i\phi)&t\cosh\theta&
i\cos\phi&0\\
0&-i\cos\phi&t\cosh\theta&-ir\cosh(\theta-i\phi)\\
i\cos\phi&0&ir\cosh(\theta+i\phi)&-t\cosh\theta
\end{pmatrix}
\end{equation}

\subsection{Contribution to $c(\phi)$ from ABS}
We shall calculate all contributions to $c$ defined by (\ref{ccc}) by inspection of all relevant excited states, depicted in Fig.\ref{abslev}.
Let us first calculate $A=\langle\psi_\pm|\partial_\phi|\psi\rangle$
where $\psi_\pm$ is one of Andreev bound states, while $\psi$ is any eigenstate.
If $\psi$ is the opposite Andreev bound state there is no contribution from spatial part $|\kappa|^{1/2}e^{-|\kappa x|}$
for the same reason as the Berry phase vanishes. If the derivative act on $e/h$ components then the contribution cancels
from orthogonality of $A_{1/2}$. The only nonzero term is when acting on $A_{1/2}$. The final result is
\begin{equation}
\langle\psi_\pm|\partial_\phi|\psi_\mp\rangle=\pm it^2r\sin^2\phi/2(1-t^2\sin^2\phi)
\end{equation}
and $c_A=t^4r^2\sin^4\phi/4(1-t^2\sin^2\phi)^2$

The relevant cross terms between ABS and scattering states are when their energies have opposite signs.
Therefore the sign $\pm$ in ABS will determine the energy (its sign is the  sign of $-\beta=\mp t\sin\phi$).
We are interested in the vector
$
\langle\psi_\pm|(|\psi_{1Ro}\rangle,|\psi_{2Ro}\rangle,|\psi_{1Lo}\rangle,|\psi_{2Lo}\rangle)\tilde\gamma
$
because $\partial_\phi$ act only on scattering matrix, $\partial_\phi S$, accompanied by outgoing states and we can insert
$S^\dag$ at the end (since the relevant quantity is Hermitian square).
The result is
\begin{eqnarray}
&&\frac{-2i|t\sin\phi|^{1/2}\sinh((\theta+i\beta)/2)}
{(2\cosh\theta)^{1/2}(|t\sin\phi|- i\mathrm{sgn}\beta\sinh\theta)}\times
\nonumber\\
&&((1\mp t\cos\phi/\cos\beta)^{1/2},\mp i(1\pm t\cos\phi/\cos\beta)^{1/2},-(1\pm t\cos\phi/\cos\beta)^{1/2},
\pm i (1\mp t\cos\phi/\cos\beta)^{1/2})\tilde\gamma
\end{eqnarray}
The final element
$
c_{B\theta}=\sum_a|\langle\pm|\partial_\phi|\psi_a\rangle|^2
$
is equal
\begin{equation}
\frac{|t^3\sin\phi|\sinh^2\theta(\cosh\theta-\cos\beta+2r^2\sin^2\phi/\cos\beta)}
{(\cosh\theta-\cos\beta)(\cosh\theta+\cos\beta)^3\cosh\theta}
\end{equation}
Its total relevant contribution is $c_B=\int_0^\infty c_{B\theta}\cosh\theta d\theta/2\pi$ which
is
\begin{equation}
\frac{|t^3\sin\phi|}{8\pi\cos^4\beta}
\left(\frac{2\beta\cos^2\beta}{\sin\beta}+2\cos^3\beta-\pi|\sin\beta|\cos^2\beta
+\cos^2\phi\left(\pi|\sin\beta|+\frac{2\beta\cos 2\beta-\sin 2\beta}{\sin^3\beta}\right)\right)
\end{equation}
For small $t$ ($\beta\to 0$) the leading terms are
$
|t^3\sin\phi|(1+2\sin^2\phi)/6\pi-t^4\sin^4\phi/8
$

\subsection{Contribution to $c(\phi)$ from scattering states}

We are interested in terms
$
c_{\theta\vartheta}=\sum_{ab}|\langle\theta_a|\partial_\phi|\vartheta_b\rangle|^2
$
where $|\theta_a\rangle$ is a scattering state with energy $-\Delta\cosh\theta$ while
$|\vartheta_b\rangle$ is a scattering state with energy $+\Delta\cosh\vartheta$. By analogous considerations
as with ABS, the derivative effectively gives $(\partial_\phi S)S^\dag=-i2\tilde\gamma$
Therefore $c$ is equal
$
4\mathrm{Tr}\tilde\gamma^2(\vartheta)Q
$
where
$
Q_{ab}=\sum_c\langle\vartheta_{oa}|\theta_c\rangle\langle\theta_c| \vartheta_{ob}\rangle
$
The middle part is
\begin{equation}
\sum_c|\theta_c\rangle\langle\theta_c|=\sum_c|\theta_{ic}\rangle\langle\theta_{ic}|+
\sum_{dc}|\theta_{od}\rangle (SS^\dag)_{dc}\langle\theta_{oc}|+
\sum_{cd}|\theta_{oc}\rangle S_{cd}\langle\theta_{id}|+\mathrm{h.c.}
\end{equation}
From unitarity of $S$ we get
\begin{equation}
\sum_c|\theta_c\rangle\langle\theta_c|=\sum_{c;x=i,o}|\theta_{xc}\rangle\langle\theta_{xc}|+
\sum_{cd}|\theta_{oc}\rangle S_{cd}\langle\theta_{id}|+\mathrm{h.c.}
\end{equation}
Therefore
$
Q=B_iB_i^\dag+B_oB_o^\dag+B_iS^\dag B_o^\dag+B_oSB_i^\dag
$
where
$
B_x=\langle\vartheta_o|\theta_x\rangle
$
with $x=i,o$.
We find
\begin{equation}
B_i=\frac{i\sinh((\vartheta-\theta)/2)}{(\cosh\theta\cosh\vartheta)^{1/2}(\sinh\theta-\sinh\vartheta)}\begin{pmatrix}
0&I\\
-I&0\end{pmatrix},\:
B_o=\frac{-i\sinh((\vartheta+\theta)/2)}{(\cosh\theta\cosh\vartheta)^{1/2}(\sinh\theta+\sinh\vartheta)}\begin{pmatrix}
I&0\\
0&-I\end{pmatrix}
\end{equation}
where $I$ is $2\times 2$ identity matrix.
Plugging in all matrices we get
\begin{equation}
c_{\theta\vartheta}=\frac{(2t\sinh\theta\sinh\vartheta)^2(\cosh\theta\cosh\vartheta+1-2\cos^2\phi-t^2\sin^2\phi)}{\cosh\theta\cosh\vartheta(\cosh\theta+\cosh\vartheta)^2(\sinh^2\theta+t^2\sin^2\phi)(\sinh^2\vartheta+t^2\sin^2\phi)}
\end{equation}
The final relevant integral is
$
c_C=\int_0^\infty d\sinh\theta d\sinh\vartheta c_{\theta\vartheta}/(2\pi)^2
$

We can divide the integral into divergent and convergent parts $c_c=c_{div}+c_{con}$
with
\begin{equation}
c_{div}=\int_0^\infty\frac{d\theta d\vartheta}{(2\pi)^2}\frac{4t^2(\cosh\theta\cosh\vartheta+1-2\cos^2\phi-t^2\sin^2\phi)}
{(\cosh\theta+\cosh\vartheta)^2}
\end{equation}
and
\begin{equation}
c_c=-\int_0^\infty\frac{d\theta d\vartheta}{(2\pi)^2}\frac{4t^4\sin^2\phi(\cosh\theta\cosh\vartheta+1-2\cos^2\phi-t^2\sin^2\phi)(t^2\sin^2\phi+\sinh^2\theta+\sinh^2\vartheta)}
{(\cosh\theta+\cosh\vartheta)^2(\sinh^2\theta+t^2\sin^2\phi)(\sinh^2\vartheta+t^2\sin^2\phi)}
\end{equation}
In the first integral we make substitution $2A=\theta+\vartheta$, $2B=\theta-\vartheta$.
The calculation needs a cutoff in the first term at $A_{max}$ and then gives
$
\pi^2c_{div}=t^2A_{max}-t^2\cos^2\phi-t^4\sin^2\phi/2
$
Now $c_{con}=c_{con1}+c_{con2}$ with
\begin{eqnarray}
&&
c_{con1}=-\int_0^\infty \frac{2t^4\sin^2\phi d\theta d\vartheta}{\pi^2(\sinh^2\vartheta+t^2\sin^2\phi)}
\frac{\cosh\theta\cosh\vartheta+1-2\cos^2\phi-t^2\sin^2\phi}{(\cosh\theta+\cosh\vartheta)^2}
\nonumber\\
&&
c_{con2}=\int_0^\infty \frac{2t^6\sin^4\phi d\theta d\vartheta
(\cosh\theta\cosh\vartheta+1-2\cos^2\phi-t^2\sin^2\phi)/\pi^2}
{(\cosh\theta+\cosh\vartheta)^2(\sinh^2\theta+t^2\sin^2\phi)(\sinh^2\vartheta+t^2\sin^2\phi)}
\end{eqnarray}

We shall integrate over $\theta$ with help of identities
\begin{eqnarray}
&&\int_0^\infty d\theta (\cosh\theta+\cosh\vartheta)^{-1}=\vartheta/\sinh\vartheta
\\
&&\int_0^\infty d\theta (\cosh\theta+\cosh\vartheta)^{-2}=\frac{\vartheta\cosh\vartheta-\sinh\vartheta}{\sinh^3\vartheta}
\nonumber\\
&&
\int_0^\infty d\theta (\cosh\theta+\cosh\vartheta)^{-3}=\frac{\vartheta(3+2\sinh^2\vartheta)-3\sinh\vartheta\cosh\vartheta}{2\sinh^5\vartheta}
\nonumber\end{eqnarray}
the integration gives
\begin{equation}
c_{con1}=-\int \frac{2d\vartheta t^4\sin^2\phi/\pi^2}{\sinh^2\vartheta+t^2\sin^2\phi}\left(\frac{\vartheta\cosh\vartheta}{\sinh\vartheta}
-(\sinh^2\vartheta+t^2\sin^2\phi+2\cos^2\phi)\frac{\vartheta\cosh\vartheta-\sinh\vartheta}{\sinh^3\vartheta}\right)
\end{equation}
We shall integrate $C_{c2}$ over $\theta$, noticing residua in $i\beta$, $i(2\pi-\beta)$, $i(\pi\pm\beta)$, and $\pm \vartheta+i\pi$.
\begin{eqnarray}
&&
c_{con2}=\int \frac{2t^6\sin^4\phi d\vartheta}{\pi^2(\sinh^2\vartheta+\sin^2\beta)}\times
\\
&&
\left[\frac{(\pi-\beta)(\cos\beta\cosh\vartheta+\cos^2\beta-2\cos^2\phi)}
{2\sin\beta\cos\beta(\cosh\vartheta+\cos\beta)^2}-\frac{\beta(\cos^2\beta-\cos\beta\cosh\vartheta-2\cos^2\phi)}
{2\sin\beta\cos\beta(\cosh\vartheta-\cos\beta)^2}\right.
\nonumber\\
&&
\left.+\frac{1-\vartheta\coth\vartheta}{\sinh^2\vartheta}
+\frac{2\cos^2\phi(1-\vartheta\coth\vartheta)-\vartheta\sinh\vartheta\cosh\vartheta}
{\sinh^2\vartheta(\sinh^2\vartheta+\sin^2\beta)}-
\frac{4\cos^2\phi\vartheta\sinh\vartheta\cosh\vartheta}
{\sinh^2\vartheta(\sinh^2\vartheta+\sin^2\beta)^2}\right]
\nonumber
\end{eqnarray}
The integration over $\vartheta$ needs noting that $\sinh^2\vartheta+\sin^2\beta=(\cosh\vartheta+\cos\beta)(\cosh\vartheta-\cos\beta)$ and further integrals  (for simplicity we assume $\beta\in[0,\pi] $)
\begin{equation}
P_1(\beta)=\int_0^\infty d\vartheta (\cosh\vartheta+\cos\beta)^{-1}=\beta/\sin\beta
\end{equation}
and by derivatives ($\sin^{-1}\beta d/d\beta$)
\begin{eqnarray}
&&P_2(\beta)=\int_0^\infty d\vartheta (\cosh\vartheta+\cos\beta)^{-2}=\frac{\sin\beta-\beta\cos\beta}{\sin^3\beta}
\\
&&
P_3(\beta)=\int_0^\infty d\vartheta (\cosh\vartheta+\cos\beta)^{-3}=\frac{\beta(3-2\sin^2\beta)-3\cos\beta\sin\beta}{2\sin^5\beta}\nonumber
\end{eqnarray}
Also
\begin{eqnarray}
&&\bar{P}_1(\beta)=\int_0^\infty d\vartheta (\sinh^2\vartheta+\sin^2\beta)^{-1}=(P_1(\pi-\beta)-P_1(\beta))/2\cos\beta
=(\pi/2-\beta)/\sin\beta\cos\beta
\\
&&
\bar{P}_2(\beta)=\int_0^\infty d\vartheta (\sinh^2\vartheta+\sin^2\beta)^{-2}=
-\bar{P}_1'(\beta)/2\sin\beta\cos\beta
\nonumber\\
&&
\bar{P}_3(\beta)=\int_0^\infty d\vartheta (\sinh^2\vartheta+\sin^2\beta)^{-2}=-
\bar{P}_2'(\beta)/4\sin\beta\cos\beta
\nonumber\end{eqnarray}

A more complicated integral
\begin{equation}
Q_1(\beta)=\int_0^\infty\frac{\vartheta\cosh\vartheta d\vartheta}{\sinh\vartheta(\cosh\vartheta+\cos\beta)}
=\frac{2\beta^2\cos\beta+\pi^2(1-\cos\beta)}{4\sin^2\beta}
\end{equation}
And its derivatives
\begin{eqnarray}
&&Q_2(\beta)=\int_0^\infty\frac{\vartheta\cosh\vartheta d\vartheta}{\sinh\vartheta(\cosh\vartheta+\cos\beta)^2}
=\frac{\pi^2(2-\sin^2\beta-2\cos\beta)+4\beta\sin\beta\cos\beta+\beta^2(2\sin^2\beta-4)}{4\sin^4\beta}
\nonumber\\
&&
Q_3(\beta)=\int_0^\infty\frac{\vartheta\cosh\vartheta d\vartheta}{\sinh\vartheta(\cosh\vartheta+\cos\beta)^3}=
\\
&&\frac{\pi^2(1-\cos\beta)+2\cos\beta\sin\beta+6\beta\sin\beta(\sin^2\beta-2)+\beta^2(2\cos\beta-\sin\beta\cos\beta+2\cos\beta\sin^2\beta)}{4\sin^6\beta}
\nonumber\\
&&
\int_0^\infty d\vartheta (\sinh^2\vartheta+\sin^2\beta)^{-1}=\frac{\pi/2-\beta}{\sin\beta\cos\beta}
\nonumber\end{eqnarray}
Another important integral is
\begin{equation}
R_1(\beta)=\int_0^\infty d\vartheta \frac{\vartheta\cosh\vartheta}{\sinh\vartheta(\sinh^2\vartheta+\sin^2\beta)}
=\int_0^\infty d\vartheta \frac{\vartheta\sinh\vartheta\cosh\vartheta}{\sin^2\beta}
[\sinh^{-2}\vartheta-
(\sinh^2\vartheta+\sin^2\beta)^{-1}]
\end{equation}
Differentiating $R_1\sin^2\beta$ over $\beta$ we get
\begin{equation}
\int_0^\infty d\vartheta \frac{2\sin\beta\cos\beta\sinh\vartheta\vartheta\cosh\vartheta}{(\sinh^2\vartheta+\sin^2\beta)^2}
\end{equation}
Integrating it by parts we get
\begin{equation}
\int_0^\infty d\vartheta \frac{\sin\beta\cos\beta}{(\sinh^2\vartheta+\sin^2\beta)^2}=(\pi/2-\beta)
\end{equation}
from the previous integral.
We note also that $R_1(\beta)\sin^2\beta\to 0$ for $\beta\to 0$ so
\begin{equation}
R_1(\beta)=\beta(\pi-\beta)/2\sin^2\beta=(Q_1(\pi-\beta)-Q_1(\beta))/2\cos\beta
\end{equation}
Its derivatives
\begin{eqnarray}
&&R_2(\beta)=\int_0^\infty d\vartheta \frac{\vartheta\cosh\vartheta}{\sinh\vartheta(\sinh^2\vartheta+\sin^2\beta)^2}
=-R_1'(\beta)/2\sin\beta\cos\beta\nonumber\\
&&R_3(\beta)=\int_0^\infty d\vartheta \frac{\vartheta\cosh\vartheta}{\sinh\vartheta(\sinh^2\vartheta+\sin^2\beta)^3}
=-R_2'(\beta)/4\sin\beta\cos\beta\end{eqnarray}

It gives $c_{con1}\pi^2/t^2$ equal
\begin{equation}
\sin^2\beta+\beta^2-\pi|\beta|+
2\cos^2\phi\left(\frac{\pi|\sin\beta-\beta\cos\beta|}{\sin^2\beta\cos\beta}
+2\cos\beta-2\beta/\sin\beta+2\frac{\beta(\beta\cos\beta-\sin\beta)}{\sin^2\beta\cos\beta}\right)
\end{equation}
and $c_{con2}\pi^2/2t^6\sin^4\phi$ equal
$$a(\beta)+b(\beta)\cos^2\phi$$
with $a$ equal
\begin{eqnarray}
&&\frac{P_1(\pi-\beta)}{2}\left(\frac{P_1(\pi-\beta)}{4\cos^2\beta}-\frac{P_2(\beta)}{2\cos\beta}
+\frac{P_1(\beta)}{4\cos^2\beta}\right)
\\
&&
+\frac{P_1(\beta)}{2}\left(\frac{P_1(\beta)}{4\cos^2\beta}+\frac{P_2(\pi-\beta)}{2\cos\beta}+\frac{P_1(\pi-\beta)}{4\cos^2\beta}\right)
-R_2(\beta)+\frac{R_1(\beta)-\bar{P}_1(\beta)-1/2}{\sin^2\beta}
\nonumber\end{eqnarray}
and $b$ equal
\begin{eqnarray}
&&-\frac{P_1(\pi-\beta)}{\cos\beta}\left(\frac{P_1(\pi-\beta)}{8\cos^3\beta}-\frac{P_3(\beta)}{2\cos\beta}-\frac{P_2(\beta)}{4\cos^2\beta}-\frac{P_1(\beta)}{8\cos^3\beta}\right)
\\
&&
+\frac{P_1(\beta)}{\cos\beta}\left(-\frac{P_1(\beta)}{8\cos^3\beta}+\frac{P_3(\pi-\beta)}{2\cos\beta}-\frac{P_2(\pi-\beta)}{4\cos^2\beta}+\frac{P_1(\pi-\beta)}{8\cos^3\beta}\right)
-4 R_3(\beta)+\frac{2R_1\beta-2\bar{P}_1(\beta)-1}{\sin^4\beta}+\frac{2R_2(\beta)-2\bar{P}_2(\beta)}{\sin^2\beta}
\nonumber
\end{eqnarray}
In the lowest order
$
c_{con}\simeq -|t^3\sin\phi|(1+2\sin^2\phi)/3\pi+2t^4\sin^4\phi/\pi^2+t^4\sin^4\phi/4
$
so the $t^3$ terms from ABS cancel!
\end{document}